# Real-space visualization of orbital-selective superconductivity in FeSe

Sang Yong Song[a,b], Gábor B. Halász[c], Jiaqiang Yan[c], Benjamin J. Lawrie[c], Petro Maksymovych*[a,d]

[a]Center for Nanophase Materials Sciences, Oak Ridge National Laboratory, Oak Ridge, TN 3781, USA

[b]Department of Physics and Chemistry, Daegu Gyeongbuk Institute of Science and Technology, Daegu, 42988, South Korea

[c]Materials Science and Technology Division, Oak Ridge National Laboratory, Oak Ridge, TN 37831, USA

[d]Department of Materials Science and Engineering, Clemson University, Clemson, SC29634, USA

*Email:  pmaksym@clemson.edu

## Abstract

We investigate the orbitally resolved superconducting properties of bulk FeSe using scanning tunneling microscopy (STM). We find that the spectral weights of both the large ($\Delta_1$) and small ($\Delta_2$) superconducting gaps remain nearly unchanged at the top Se sites as the STM tip approaches atomic contact. In contrast, the spectral weight of $\Delta_2$ increases significantly at the Fe and bottom Se sites. These results suggest that the gap $\Delta_2$ is localized in the *xy*-plane and likely associated with the $d_{xy}$ orbital band. Furthermore, we observe a long-range suppression of the large gap $\Delta_1$ near one-dimensional (1D) defects such as twin boundaries, wrinkles, and step edges, whereas $\Delta_2$ remains robust. This indicates that the two superconducting gaps respond differently to such 1D defects. High-resolution measurement using a Pb-coated tip reveals localized in-gap states near 1D defects, indicating possible defect-induced magnetism. Our findings highlight the contrasting behaviors of gap $\Delta_1$ and gap $\Delta_2$ in response to local electronic and magnetic environments and provide real-space evidence for orbital-selective superconductivity.

## Introduction

The multiple Fe 3d orbitals in Fe-based superconductors play a central role in spin fluctuations, nematicity, and Cooper pairing. In such multi-orbital systems, strong Hund's coupling has been proposed to equalize orbital populations and suppress inter-orbital charge fluctuations, leading to orbital decoupling[1,2]. As a result, quasiparticles from different orbital bands acquire orbital-dependent renormalized weights—a hallmark of Hund metals[3-6]. Within this framework, both Cooper pairing[7,8] and magnetic fluctuations[9-11] are expected to exhibit orbital selectivity. Depending on the strength of the onsite Coulomb interaction and Hund's coupling, electrons in the $d_{xz}/d_{yz}$ orbitals remain itinerant, while those in the $d_{xy}$ orbital can become localized in an orbital-selective Mott phase[12,13]. This orbital-selective picture successfully accounts for the observed electronic structure[14], spin fluctuations[15-20], nematic order[20,21], and superconducting gap anisotropies[22-25] in Fe-based superconductors.

Sprau et al.[22] provided experimental evidence for orbital-selective Cooper pairing in FeSe through Bogoliubov quasiparticle interference (BQPI) measurements. They identified two anisotropic superconducting gaps, $\Delta_1$ (2.3 mV) and $\Delta_2$ (1.5 mV), associated with a hole pocket composed of $d_{yz}$ and $d_{xz}$ orbitals at the $\Gamma$ point and an electron pocket composed of $d_{yz}$ and $d_{xy}$ orbitals at the X point. Their analysis suggests that Cooper pairing is primarily mediated by the $d_{yz}$ orbital. Tong Chen et al.[17] further demonstrated strong anisotropic spin excitations at low energies (5-10 meV) in detwinned FeSe using inelastic neutron scattering. By resolving the orbital character of spin susceptibility, they concluded that the observed anisotropy arises from orbital-selective spin excitations dominated by the $d_{yz}$ orbital. Resonant inelastic X-ray scattering measurements[20] revealed that such anisotropic spin excitations persist up to much higher energies (~200 meV), well above the $d_{yz}/d_{xz}$ orbital splitting (~50 meV). This observation supports the view that nematic order in FeSe is driven by magnetic rather than orbital fluctuations. Moreover, the dispersive and underdamped nature of high-energy anisotropic spin excitations is well captured by theoretical models incorporating local moment physics, such as the Hund's metal picture[26], the orbital-selective Mott phase[27], and the Mott–Hund's scenario[28]. In parallel, such anisotropic features can also be accounted for by itinerant approaches that incorporate strong orbital-selective renormalizations and damping[29].

Unlike many other Fe-based superconductors, FeSe does not exhibit long-range magnetic order at ambient pressure[30-32], suggesting that its magnetism is frustrated and that the system lies close to a magnetic quantum critical point[33,34]. Several studies have shown that local perturbations–including vacancies[35-38], non-magnetic impurities[36,37,39,40], nematic twin boundaries[38], and atomic steps[36,37]–can induce local magnetic moments in FeSe. Notably, the induced magnetic patterns around defects display $C_2$ symmetry, consistent with predictions based on orbital-selectivity[35]. Moreover, theoretical calculations suggest that a 100-type step edge generates a sharp peak in local magnetization, with a finite magnetic tail extending into the surrounding lattice[36].

In this study, we apply proximity tunneling in scanning tunneling microscopy (STM) to investigate orbital-resolved superconductivity in FeSe in real-space. We find that the spectral weights of the two superconducting gaps depend on atomic sites near the atomic contact regime. Based on our observations, we conclude that the small gap, $\Delta_2$, is confined to the *xy*-plane or exhibits in-plane momentum characteristics. This suggests that the large gap $\Delta_1$ is dominated by the $d_{yz}$ orbital[8,22], while $\Delta_2$ is associated with the $d_{xy}$ orbital and exhibits orbital-selective characteristics.

Notably, the spectral weight of the gap $\Delta_1$ is suppressed over extended distances near one-dimensional defects such as twin boundaries, wrinkles, and atomic step edges, whereas gap $\Delta_2$ remains largely unaffected. High-resolution spectroscopy using a Pb-coated superconducting tip reveals additional in-gap states emerging at the centers of these one-dimensional (1D) defects, appearing at energies below the gap $\Delta_2$. These states are reminiscent of magnetic bound states arising from locally induced magnetic moments. We suggest that the independent behavior of the two superconducting gaps near 1D-defects is a consequence of orbital-selective superconductivity.

**Results**

FeSe hosts a hole pocket at $\Gamma$ and an electron pocket at X at the Fermi level, primarily composed of Fe $d_{xz}$, $d_{yz}$, and $d_{xy}$ orbitals, as established by angle-resolved photoemission spectroscopy (ARPES)[41-43], quantum oscillation (QO)[44,45], and quasiparticle interference (QPI)[46] studies. In this multi-orbital system, strong Hund's coupling leads to orbital-dependent quasiparticle weights, which are expected to influence the superconducting pairing. Notably, Bogoliubov quasiparticle interference (BQPI) measurements have shown that Cooper pairing is primarily mediated by the $d_{yz}$ orbital[22].

FeSe has a layered structure consisting of a planar square lattice of iron (Fe) atoms sandwiched between two layers of selenium (Se) atoms (Fig. 1a). Figure 1b shows a top view of the atomic structure, where brown circles represent Fe atoms. Green circles with a thin black outline and solid green circles indicate the top and bottom Se atoms, respectively. In the STM image, only the Se atoms in the top layer are visible as bright spots (inset of Fig. 1c).

To probe orbital-resolved superconductivity in real space, we gradually approached the FeSe surface with the STM tip to enter the atomic contact regime, and investigated how the spectral weights of the two superconducting gaps vary at different atomic sites. Figure 1c shows the conductance curves on a logarithmic scale as the STM tip approaches each atomic site, starting from the set point of 5 mV and 1 nA (defined as z = 0 pm). At the top Se site (green circles with a thin black outline in the inset), the conductance increases logarithmically as the tip approaches, up to z = -200 pm (black circles in Fig. 1c). Both the large gap ($\Delta_1$) and the small gap ($\Delta_2$) remain nearly unchanged regardless of the tip's distance z (Fig. 1d). It is likely that the interaction between the STM tip and the Fe d orbitals is impeded by the top Se atom (Fig. 1g). In contrast, at the Fe site (solid brown circle in the inset of Fig. 1c), a sudden increase in conductance is observed around z = -160 pm (brown circles in Fig. 1c), indicating entry into the near-atomic-contact regime[47]. This behavior implies that the tip–sample interaction at the Fe site is qualitatively different from that at the top Se site. Notably, from z = -160 pm, the spectral weight of gap $\Delta_2$ increases significantly (Fig. 1e), suggesting enhanced overlap with an orbital channel involved in the formation of $\Delta_2$ (Fig. 1h). At the bottom Se site (solid green circle in the inset of Fig. 1c), the conductance curve exhibits two step-like increases, first at z = -130 pm and then around -160 pm (green circles in Fig. 1c). As shown in Fig. 1f, the spectral weight of gap $\Delta_2$ increases at z = -140 pm, and remarkably, at z = -160 pm, it exceeds that of gap $\Delta_1$. This indicates that as the STM tip approaches the bottom Se site, the interaction between the tip and the orbital channel associated with $\Delta_2$ becomes stronger than at the Fe site (Fig. 1i), thereby enhancing the contribution of $\Delta_2$ (Fig. 1f). These results suggest that the superconducting gap $\Delta_2$ likely originates from the $d_{xy}$ orbital channel, which lies in the *xy*-plane.

To further investigate the spatial variation of spectral weights of the two superconducting gaps at atomic scale, we measured atomic-resolution dI/dV maps near the atomic contact regime (set point: $V_{bias}$ = -4.5 mV, I = 21 nA, 0.06 $G_0$). Figure 2a shows a topographic image of FeSe acquired during the dI/dV mapping. The black circles indicate top Se atomic sites. The dI/dV spectra shown in Figures 2e and 2g were acquired at the top Se site (blue arrow in Fig. 2a) and the bottom Se site (red arrow in Fig. 2a) under two conditions: the near-atomic-contact regime (0.06 $G_0$) and the tunneling regime ($V_{bias}$ = -4.5 mV, I = 1 nA, 0.003 $G_0$), respectively. In the near-contact regime, the spectral weight of the small gap $\Delta_2$ increases noticeably at the bottom Se sites (gray arrow in Fig. 2e), whereas in the tunneling regime, the spectral weight remains unchanged (Fig. 2g). Figure 2d shows how the superconducting gap structure varies along several top and bottom Se atomic sites (red dotted line in Fig. 2a) in the near-atomic-contact regime. The large gap $\Delta_1$ is dominant at the top Se sites (black dashed lines), while the intensity of small gap $\Delta_2$ shows a pronounced enhancement at the bottom Se sites (gray dashed lines). In contrast, under typical tunneling conditions ($V_{bias}$ = -4.5 mV, I = 1 nA, 0.003 $G_0$), the superconducting gaps show minimal variation along the same path (Fig. 2f). In the near-atomic-contact regime, the differential conductance map at the energy corresponding to gap $\Delta_1$ (~2 meV) (Fig. 2b) shows that the spectral weight of $\Delta_1$ is primarily localized at the top Se sites. In contrast, the differential conductance map at the energy corresponding to gap $\Delta_2$ (~1 meV) (Fig. 2c) reveals that $\Delta_2$ is mainly distributed at the Fe and bottom Se sites. We propose that these observations provide real-space evidence for orbital-resolved superconductivity, and that $\Delta_1$ and $\Delta_2$ are dominated by the contributions of the $d_{yz}$ and $d_{xy}$ orbitals, respectively.

In addition, we observed changes in the spectral weights of gaps $\Delta_1$ and $\Delta_2$ near one-dimensional (1D) defects such as wrinkles, twin boundaries, and step edges in FeSe. Figure 3b shows how the spectral weights evolve near a wrinkle (red dots in Fig. 3a), with noticeable changes beginning approximately 40 nm from its center. As the spectral weight of $\Delta_1$ gradually diminishes approaching the center, only $\Delta_2$ remains (Fig. 3b). Interestingly, a twin boundary (Fig. 3c) exhibits a similar behavior: suppression of $\Delta_1$ near the center and a relative enhancement of $\Delta_2$ (Fig. 3d), with the transition starting around 10 nm away. Likewise, the step edge (Fig. 3e) alters the spectral weights of both gaps starting ~10 nm from the edge (Fig. 3f). Notably, the step edge induces distinct in-gap states (dark green spectra in Fig. 3f and red spectra in Fig. S1), reminiscent of Yu–Shiba–Rusinov states[48-50]. Altogether, these results show that gaps $\Delta_1$ and $\Delta_2$ respond differently to 1D defects.

To investigate the detailed gap structures around 1D defects in FeSe, we employed a Pb-coated superconducting tip (Fig. 4). The superconducting gap size of the Pb-coated tip ($\Delta_{tip}$) was determined to be ±1.37 meV based on the energy position of the Andreev resonance peaks (red arrows in Fig. S2b). The use of a superconducting tip causes the measured coherence peaks to shift by $\pm\Delta_{tip}$ due to the convolution of the sample and tip superconducting gaps[51]. From spectra acquired using both normal and superconducting tips (Figs. 3, 4, and S2), we determined that $\Delta_1$ consists of at least three peaks (black arrows in Fig. S2), and we additionally observed another coherence peak, $\Delta_3$, at ±0.6–0.8 meV (corresponding to $\pm(\Delta_{tip} + \Delta_3) \approx \pm2.04$–2.19 meV; green arrows in Fig. S2). These features are not resolvable with the normal tip at 1.2 K.

Figure 4a shows a differential conductance map at 3 mV across one-unit-cell (1UC) and two-unit-cell (2UC) step edges of FeSe (see also Fig. S3). A twin boundary is located near the left edge of the 1UC terrace (blue arrow), and a superconducting vortex is visible on the twin boundary and another on the terrace (yellow arrows). Figure 4d displays the spatial evolution of superconducting gaps along the black dotted line in Fig. 4a. On the terrace, gap $\Delta_1$ varies from ±1.31 to ±2.67 meV (i.e., $\pm(\Delta_{tip} + \Delta_1) \approx \pm2.67$-4.04 meV; gray boxes in Fig. 4b and Fig. S4a), gap $\Delta_2$ ranges from ±0.96 to ±1.36 meV (±2.34–2.74 meV; blue boxes), and gap $\Delta_3$ appears at ±0.66 to ±0.81 meV (±2.04–2.19 meV; green boxes). The estimated energy ranges of all coherence peaks are summarized in Supplementary Table S1. The superconducting gaps are similarly changed near both the twin boundary (Fig. 4e) and the step edges (Fig. 4f and 4g). The black, sky blue, and blue spectra in Fig. 4b represent the superconducting gaps on the terrace, near the twin boundary, and at the center of the twin boundary, respectively. The gray, blue, and green boxes indicate the energy ranges of gaps $\Delta_1$, $\Delta_2$, and $\Delta_3$, respectively. (Note that all coherence peaks are shifted by $\Delta_{tip}$ (±1.37 meV) due to the use of the Pb-coated tip.) Near the twin boundary, the spectral weight of $\Delta_1$ is strongly suppressed (black arrows in Fig. 4b), while that of $\Delta_2$ is relatively enhanced (sky blue arrows in Fig. 4b). Remarkably, at the center of the twin boundary (blue spectrum in Fig. 4b), $\Delta_1$ is suppressed and a new set of peaks emerges within the $\Delta_3$ (blue arrows in Fig. 4b and Fig. 4e). Near a step edge (orange spectrum in Fig. 4c), $\Delta_1$ is suppressed (black arrows), while $\Delta_2$ is enhanced (orange arrows), showing similar behavior to the twin boundary. At the 1-UC and 2-UC step edges, the green and red spectra (Fig. 4c) reveal additional peaks at energies lower than $\Delta_3$ (green and red arrows in Fig. 4c, 4f, and 4g), which are discussed in more detail below.

We estimated the detailed structure of the superconducting gaps by tracking changes in the energy positions of the coherence peaks near the terrace, twin boundary, and step edges (Fig. S4).

Figure 5 presents the spatial variation of the dI/dV spectra and the corresponding trajectories of the gaps $\Delta_1$, $\Delta_2$, $\Delta_3$, and in-gap states on the terrace (Figs. 5a and 5b), near the twin boundary (Figs. 5c and 5d), and near the 2-UC step edge (Figs. 5e and 5f). Gap $\Delta_1$ consists of at least three distinct peaks, which oscillate on the terrace within the range of 1.31–2.67 mV ($\Delta_1 + \Delta_{tip}$ = 2.67–4.04 meV) (black circles in Fig. 5b and Fig. S5b). Due to the relatively low spectral weight of $\Delta_2$ and $\Delta_3$ on the terrace, the presence and exact positions of $\Delta_3$ are sometimes difficult to determine. The estimated energy ranges of $\Delta_2$ and $\Delta_3$ are approximately 0.96–1.36 meV ($\Delta_2 + \Delta_{tip}$ = 2.34–2.74 meV) (blue circles in Fig. 5b and Fig. S5b) and 0.66–0.81 meV ($\Delta_3 + \Delta_{tip}$ = 2.04–2.19 meV) (green circles in Fig. 5b and Fig. S5b), respectively. The origin of the multiple gaps is not yet fully understood, but it is likely associated with fine splitting in the Fermi pockets, as revealed by high-resolution, laser-based angle-resolved photoemission spectroscopy (ARPES) measurements[52].

Figures 5c and 5e show the spatial variation of superconducting gaps near the twin boundary and the step edge, respectively. In both cases, a set of new peaks appears at the center (x = 0) of the twin boundary and the step edge, with tails of in-gap states extending into the terrace by approximately 5–8 nm (red circles in Figs. 5d and 5f, and red triangles in Figs. S4b and S4c). In addition, the spectral weight of all $\Delta_1$ peaks begins to diminish around 15 nm from the twin boundary, with two of the peaks either merging or disappearing completely (black circles in Figs. 5d and 5f, and black triangles in Figs. S4b and S4c). The spectral weight of gap $\Delta_2$ becomes relatively enhanced, and the energy of gap $\Delta_2$ slightly decreases near the twin boundary and step edge (blue circles in Figs. 5d and 5f, and blue triangles in Figs. S4b and S4c). Although the wrinkle has a less clearly defined boundary and a broader affected region (Fig. 3a and Fig. S6), the spectral weight of gap $\Delta_1$ also decreases, while that of gap $\Delta_2$ is relatively enhanced, and in-gap states appear at the center—similar to what is observed at the twin boundary and step edge (Fig. S6). These results indicate that the twin boundary, step edge, and wrinkle host localized in-gap states at their centers, and that $\Delta_1$ and $\Delta_2$ exhibit universal spatial variations near these 1D defects (see direct comparison in Fig. S7). The spatial distribution of in-gap states along the step edge is irregular and localized (Fig. S8), likely due to the atomically irregular nature of the FeSe step edges (Figs. 3e and S8a) on our sample.

**Discussion**

By approaching the FeSe surface with the STM tip, we enhanced the interaction between the tip and the Fe orbital channels, enabling direct measurement of the contributions of the superconducting gaps $\Delta_1$ and $\Delta_2$ from individual orbital channels at distinct atomic sites (Fig. 1). In the near-atomic-contact regime, we confirmed that the gap $\Delta_2$ is associated with the $d_{xy}$ orbital channel, as evidenced by a significant increase in its spectral weight when the STM tip approaches the Fe and bottom Se sites, where the $d_{xy}$ orbital contribution is maximized. Furthermore, we confirmed that the spectral weight of gap $\Delta_1$ is gradually suppressed starting from a relatively long distance (15 ~ 40 nm) away from 1D defects such as wrinkles, twin boundaries, and step edges, while gap $\Delta_2$ remains largely unaffected near these defects (Figs. 3 and 4). Using a Pb-coated superconducting tip, we detected in-gap states near these 1D defects.

Interestingly, the peak of the in-gap state appears to emerge from the atomic scale, and the distribution (Fig. S8) is reminiscent of Yu-Shiba-Rusinov states arising from proximity-induced s-wave Cooper pairs, triggered by a local magnetic moment at the step edge of FeSe films grown on Pb(111)[37]. In addition, the theoretical prediction in Ref. 36—that FeSe step edges can induce local magnetization and generate a finite magnetization tail extending into the bulk—aligns well with our observations of local in-gap states and the spatial variation of the superconducting gap $\Delta_1$. Taken together, we suggest that the in-gap states observed here may originate from locally induced magnetism at twin boundaries, wrinkles, and step edges. Furthermore, if magnetic fluctuations associated with each orbital band are orbital-selective and respond differently to these 1D defects, the corresponding orbital-selective superconductivity may also be differently affected. In particular, the relatively localized nature of electrons in the $d_{xy}$ band may lead to a shorter-range response to 1D defects. This can explain why the $\Delta_1$ gap begins to be suppressed at relatively long distances from 1D defects, while the $\Delta_2$ gap remains largely unaffected. Nevertheless, the effect of the nontrivial phase structure of a sign-changing superconducting order parameter[54–56] cannot be excluded as a possible contributing factor.

Previous studies have reported that Fe-based superconductors can exhibit either single or multiple superconducting gaps depending on sample orientation and disorder[23, 57-61], both of which can be a result of orbital-selective magnetism and superconductivity. It remains to be theoretically verified whether the interpretation that the $\Delta_2$ gap is confined to the *xy*-plane is consistent with the observed behavior. A deeper understanding will also require theoretical analysis of how different orbital channels contribute to magnetism near one-dimensional defects and influence superconductivity. Our findings provide new insights into the real-space manifestation of orbital-selective superconductivity in Fe-based superconductors.

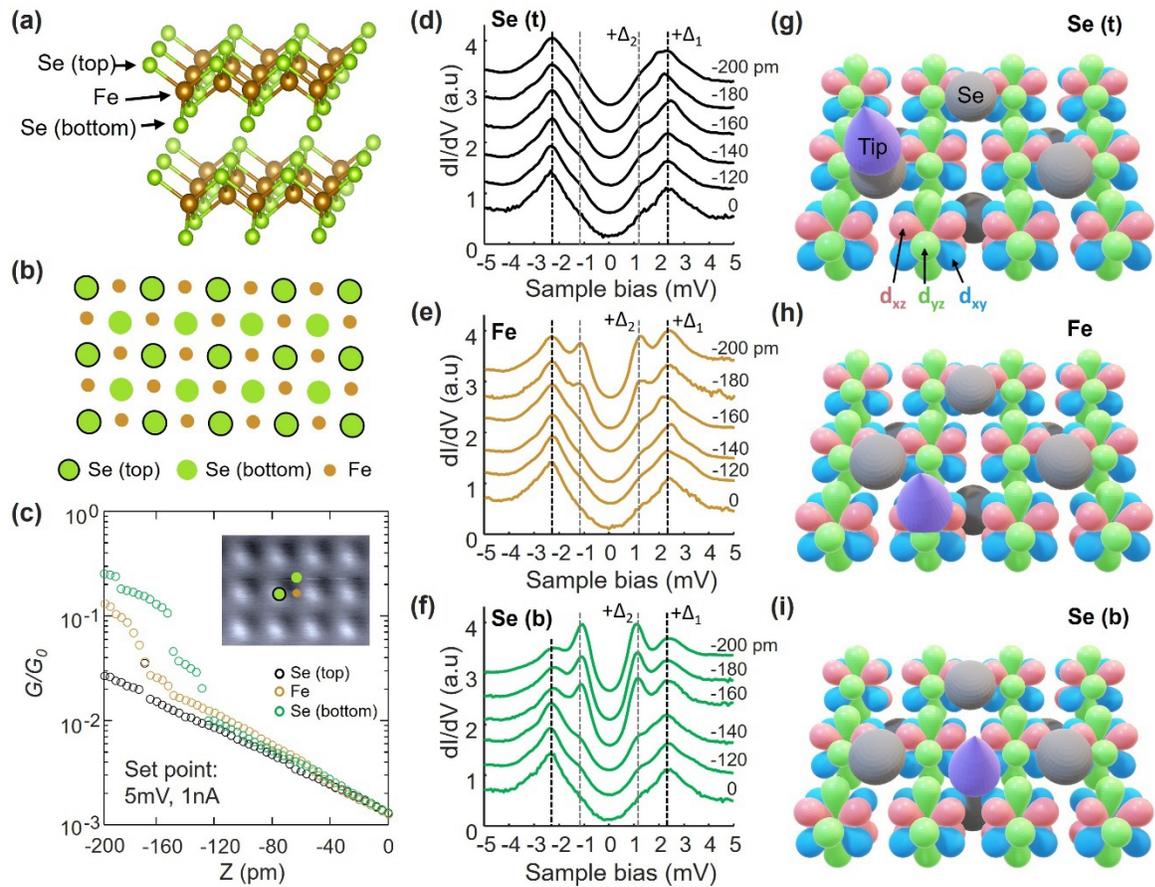

**Figure 1 Near atomic contact on FeSe and atomic resolved dI/dV spectra. (a)** Schematic image of atomic structure of FeSe. **(b)** The top view of the atomic structure of FeSe. **(c)** Conductance curves as the tip approaches each atomic sites from the set point of 5 mV 1 nA (z = 0 pm) to -200 pm. Black, brown, and green conductance curves are obtained at the top Se site (green circle with thin black line in inset), the Fe site (brown circle in inset), and the bottom Se site (green solid circle in inset), respectively. **(d-f)** dI/dV spectra as the tip approaches at (d) the top Se site, (e) the Fe site, and (f) the bottom Se site. **(g-i)** Schematic images of overlap between Fe d-channels ($d_{xz}$: red, $d_{yz}$: green, $d_{xy}$: blue) and tip (purple) at (g) the top Se site, (h) the Fe site, and (i) the bottom Se site.

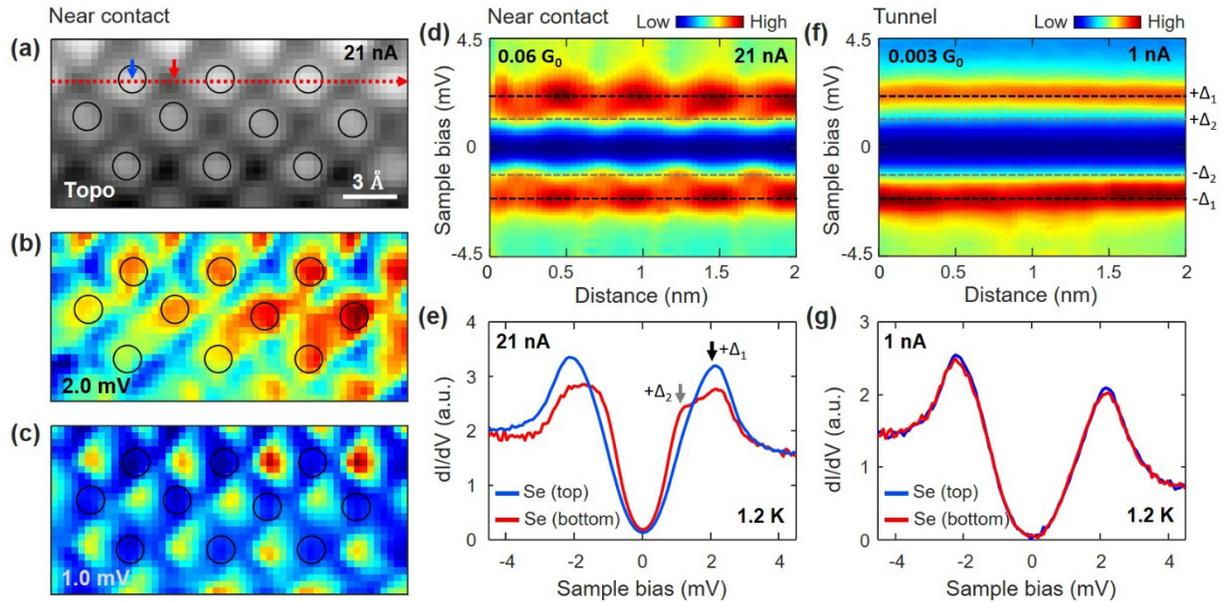

**Figure 2 Atomic-resolved differential conductance maps. (a)** Atomic resolution STM image of FeSe (Setpoint: $V_{bias}$ = 4.5 mV, I = 21 nA). Black circles indicate the positions of top Se atoms. **(b, c)** Differential conductance maps at (b) 2.0 mV and (c) 1.0 mV near the atomic contact (0.06 $G_0$). **(d)** Spatial variation of superconducting gaps across the red line in (a) near the atomic contact (0.06 $G_0$). The contribution of gap $\Delta_2$ increases at the Fe site. **(e)** dI/dV spectra obtained at the top Se site (blue), and the Fe site (red) near atomic contact (0.06 $G_0$). **(f)** Spatial variation of superconducting gap across the red line in (a) in tunneling regime (0.003 $G_0$). **(g)** dI/dV spectra obtained at the top Se site (blue), and the Fe site (red) in tunneling regime (0.003 $G_0$).

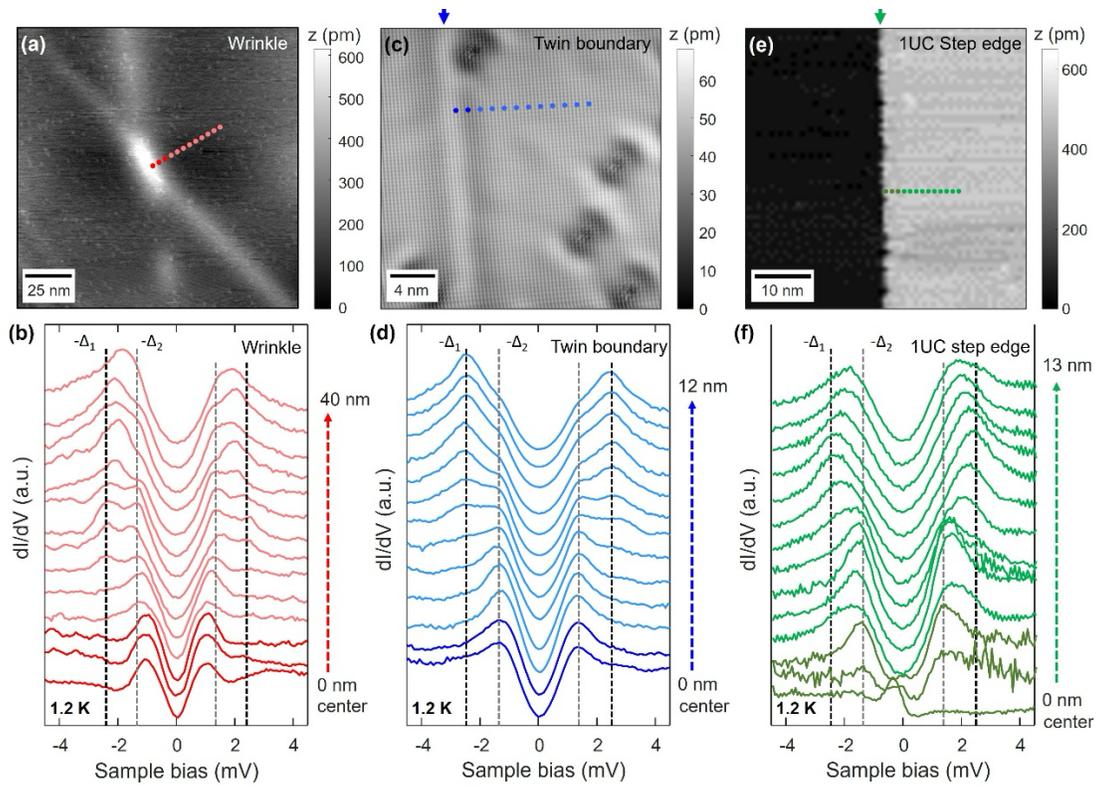

**Figure 3 Spatial variation of superconducting gap near a wrinkle, a twin boundary, and a step edge of FeSe (a)** STM image of wrinkle on FeSe surface ($V_{bias}$ = 10 mV, I = 100pA). **(b)** Spatial variation of superconducting gap along the red dotted line in (a). **(c)** STM image of the twin boundary (blue arrow). **(d)** Spatial variation of superconducting gap near the twin boundary (blue dotted line in (c)). **(e)** STM image of the one-unit cell step of FeSe. **(f)** Spatial variation of superconducting gap near step edge (green dotted line).

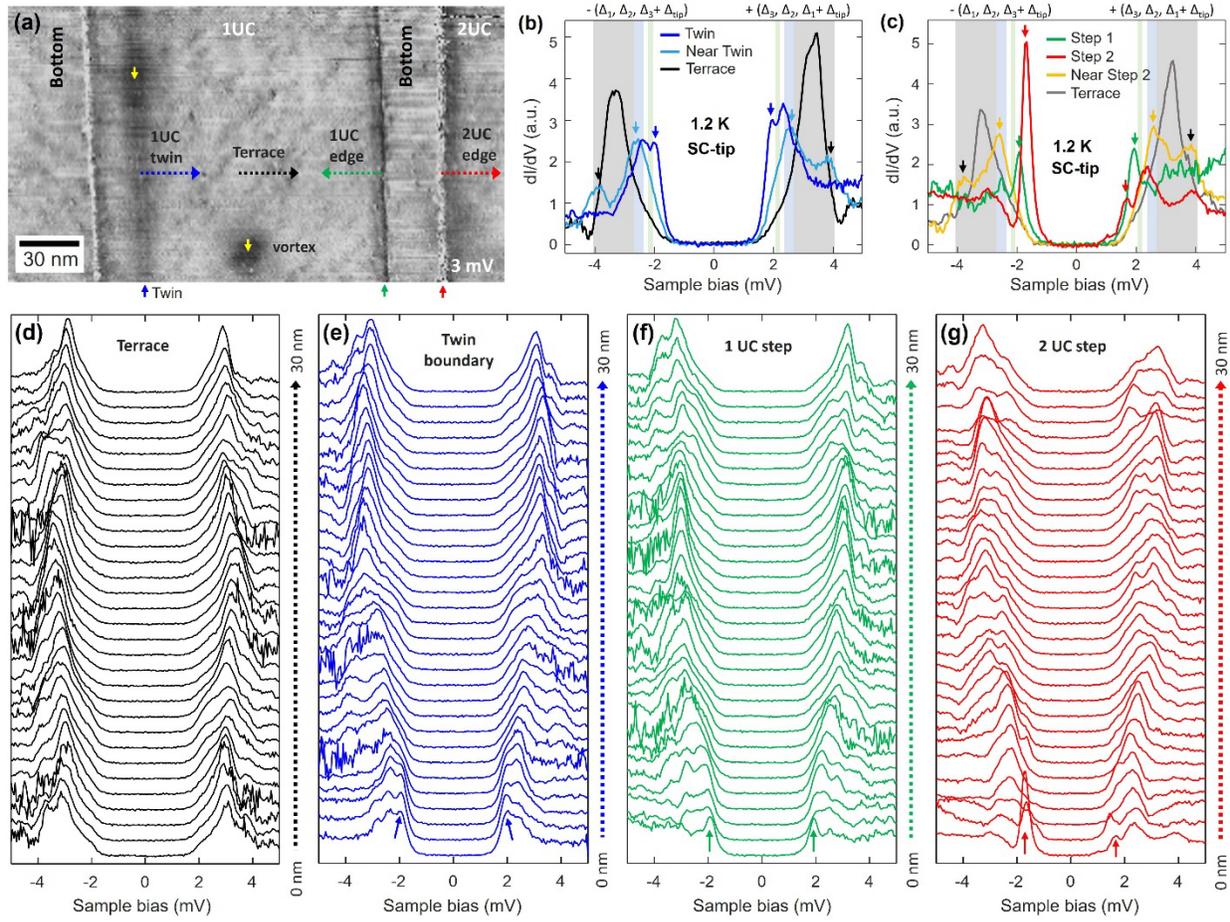

**Figure 4 High energy resolution spectra near a twin boundary and step edges of FeSe. (a)** Differential conductance map of steps (green and red arrows) and the twin boundary (blue arrow) at 3 mV. **(b)** The dI/dV spectra obtained on the twin (blue), near twin (sky blue), and the terrace (black) using a Pb-coated superconducting tip. The gray, blue, and green boxes indicate the energy ranges of gap $\Delta_1$, $\Delta_2$, and $\Delta_3$, respectively. A set of in-gap state appears on the twin boundary (blue arrows). **(c)** The dI/dV spectra obtained on the step 1 (green), on the step 2 (red), near the step 2 (orange), and on the terrace (gray). The in-gap states appear at the step 1 (green arrows), and the step2 (red arrows). **(d - g)** Spatial variations of superconducting gaps across the terrace, the twin boundary, and steps (black, blue, green, red dashed arrows in (a)) obtained using a Pb-coated superconducting tip at 1.2 K.

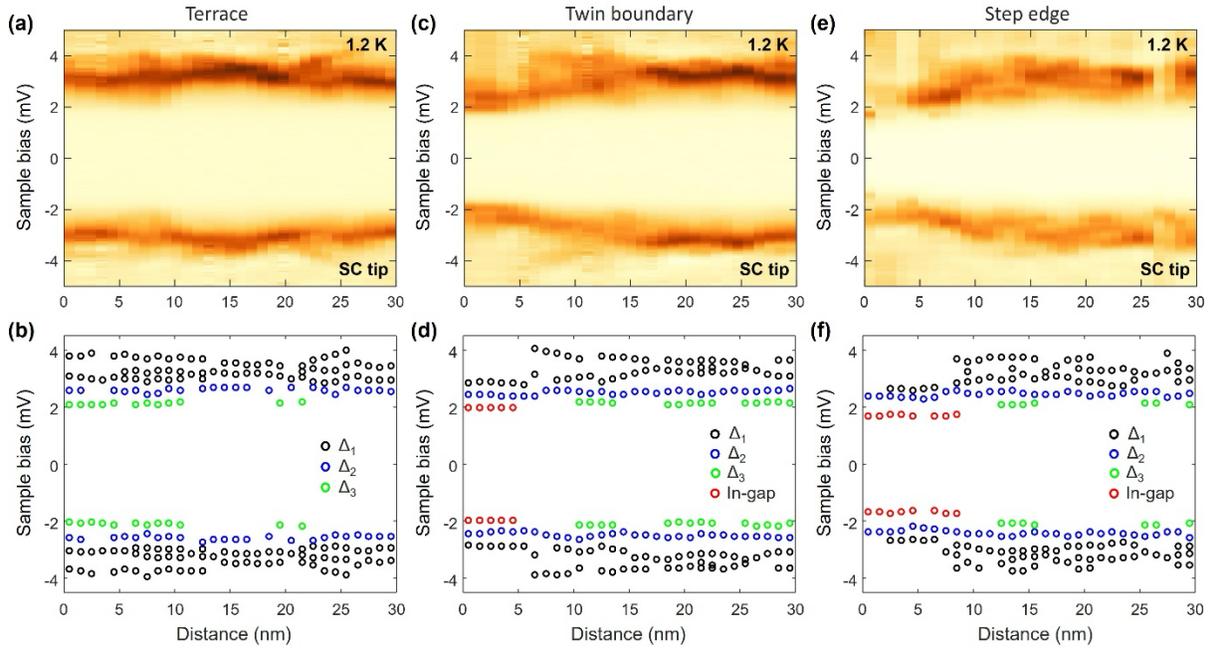

**Figure 5 The trajectory of the superconducting coherence peaks on the terrace, near twin boundary, and step edge. (a)** The spatial variation of superconducting gaps on the terrace measured by Pb-coated superconducting tip. **(b)** The trajectory of the superconducting coherence peaks of the gap $\Delta_1$ (black circles), $\Delta_2$ (blue circles), and $\Delta_3$ (green circles). The gap $\Delta_1$ consists of three branches. **(c, e)** The spatial variation of superconducting gaps near twin boundary and step edge, respectively. **(d, f)** The trajectories of the superconducting coherence peaks of the gap $\Delta_1$ (black circles), $\Delta_2$ (blue circles), $\Delta_3$ (green circles), and in-gap states (red circles) near twin boundary and step edge, respectively.

# References

1. L. De' Medici, S. R. Hassan, M. Capone, X. Dai, Orbital-Selective Mott Transition out of Band Degeneracy Lifting. *Phys. Rev. Lett.* **102**, 126401 (2009).

2. L. De' Medici, Hund's coupling and its key role in tuning multiorbital correlations. *Phys. Rev. B* **83**, 205112 (2011).

3. Z. P. Yin, K. Haule, G. Kotliar, Kinetic frustration and the nature of the magnetic and paramagnetic states in iron pnictides and iron chalcogenides. *Nature Mater* **10**, 932–935 (2011).

4. N. Lanatà, H. U. R. Strand, G. Giovannetti, B. Hellsing, L. De' Medici, M. Capone, Orbital selectivity in Hund's metals: The iron chalcogenides. *Phys. Rev. B* **87**, 045122 (2013).

5. A. Georges, L. de' Medici, J. Mravlje, Strong correlations from Hund's coupling. *Annual Review of Condensed Matter Physics* **4**, 137–178 (2013).

6. L. De' Medici, G. Giovannetti, M. Capone, Selective Mott Physics as a key to iron superconductors. *Phys. Rev. Lett.* **112**, 177001 (2014).

7. R. Yu, J.-X. Zhu, Q. Si, Orbital-selective superconductivity, gap anisotropy, and spin resonance excitations in a multiorbital t - J 1 - J 2 model for iron pnictides. *Phys. Rev. B* **89**, 024509 (2014).

8. A. Kreisel, B. M. Andersen, P. O. Sprau, A. Kostin, J. C. S. Davis, P. J. Hirschfeld, Orbital selective pairing and gap structures of iron-based superconductors. *Phys. Rev. B* **95**, 174504 (2017).

9. H. Lee, Y.-Z. Zhang, H. O. Jeschke, R. Valentí, Orbital-selective phase transition induced by different magnetic states: A dynamical cluster approximation study. *Phys. Rev. B* **84**, 020401 (2011).

10. L. Fanfarillo, J. Mansart, P. Toulemonde, H. Cercellier, P. L. Fevre, F. Bertran, B. Valenzuela, L. Benfatto, V. Brouet, Orbital-dependent Fermi surface shrinking as a fingerprint of nematicity in FeSe. *Phy. Rev. B* **94**, 155138 (2016).

11. L. Fanfarillo, L. Benfatto, B. Valenzuela, Orbital mismatch boosting nematic instability in iron-based superconductors. *Phys. Rev. B* **97**, 121109 (2018).

12. M. Yi, D. H. Lu, R. Yu, S. C. Riggs, J.-H. Chu, B. Lv, Z. K. Liu, M. Lu, Y.-T. Cui, M. Hashimoto, S.-K. Mo, Z. Hussain, C. W. Chu, I. R. Fisher, Q. Si, Z.-X. Shen, Observation of temperature-induced crossover to an orbital-selective Mott phase in $A_xFe_{2-y}Se_2$ (A = K, Rb) Superconductors. *Phys. Rev. Lett.* **110**, 067003 (2013).

13. R. Yu, H. Hu, E. M. Nica, J. X. Zhu, Q. Si, Orbital selectivity in electron correlations and superconducting pairing of iron-based superconductors. *Frontiers in Physics*, **9**, 578347 (2021).

14. M. Yi, Y. Zhang, Z.-X. Shen, D. Lu, Role of the orbital degree of freedom in iron-based superconductors. *npj Quant Mater* **2**, 1–12 (2017).

15. Y. Li, Z. Yin, X. Wang, D. W. Tam, D. L. Abernathy, A. Podlesnyak, C. Zhang, M. Wang, L. Xing, C. Jin, K. Haule, G. Kotliar, T. A. Maier, P. Dai, Orbital selective spin excitations and their Impact on superconductivity of $LiFe_{1-x}Co_xAs$. *Phys. Rev. Lett.* **116**, 247001 (2016).

16. J.-H. She, M. J. Lawler, E.-A. Kim, Quantum spin liquid intertwining nematic and superconducting order in Fese. *Phys. Rev. Lett.* **121**, 237002 (2018).

17. T. Chen, Y. Chen, A. Kreisel, X. Lu, A. Schneidewind, Y. Qiu, J. T. Park, T. G. Perring, J. R. Stewart, H. Cao, R. Zhang, Y. Li, Y. Rong, Y. Wei, B. M. Andersen, P. J. Hirschfeld, C. Broholm, P. Dai, Anisotropic spin fluctuations in detwinned FeSe. *Nature Mater* **18**, 709–716 (2019).

18. D. W. Tam, Z. Yin, Y. Xie, W. Wang, M. B. Stone, D. T. Adroja, H. C. Walker, M. Yi, P. Dai, Orbital selective spin waves in detwinned NaFeAs. *Phys. Rev. B* **102**, 054430 (2020).

19. H. Miao, Y. L. Wang, J.-X. Yin, J. Zhang, S. Zhang, M. Z. Hasan, R. Yang, X. C. Wang, C. Q. Jin, T. Qian, H. Ding, H.-N. Lee, G. Kotliar, Hund's superconductor Li(Fe,Co)As. *Phys. Rev. B* **103**, 054503 (2021).

20. X. Lu, W. Zhang, Y. Tseng, R. Liu, Z. Tao, E. Paris, P. Liu, T. Chen, V. N. Strocov, Y. Song, R. Yu, Q. Si, P. Dai, T. Schmitt, Spin-excitation anisotropy in the nematic state of detwinned FeSe. *Nature Phys* **18**, 806–812 (2022).

21. S.-H. Baek, D. V. Efremov, J. M. Ok, J. S. Kim, J. van den Brink, B. Büchner, Orbital-driven nematicity in FeSe. *Nature Mater* **14**, 210–214 (2015).

22. P. O. Sprau, A. Kostin, A. Kreisel, A. E. Böhmer, V. Taufour, P. C. Canfield, S. Mukherjee, P. J. Hirschfeld, B. M. Andersen, J. C. S. Davis, Discovery of orbital-selective Cooper pairing in FeSe. *Science* **357**, 75–80 (2017).

23. J.-X. Yin, X.-X. Wu, J. Li, Z. Wu, J.-H. Wang, C.-S. Ting, P.-H. Hor, X. J. Liang, C. L. Zhang, P. C. Dai, X. C. Wang, C. Q. Jin, G. F. Chen, J. P. Hu, Z.-Q. Wang, A. Li, H. Ding, S. H. Pan, Orbital selectivity of layer-resolved tunneling in the iron-based superconductor Ba 0.6 K 0.4 Fe 2 As 2. *Phys. Rev. B* **102**, 054515 (2020).

24. S. Bhattacharyya, P. J. Hirschfeld, T. A. Maier, D. J. Scalapino, Effects of momentum-dependent quasiparticle renormalization on the gap structure of iron-based superconductors. *Phys. Rev. B* **101**, 174509 (2020).

25. C. Liu, A. Kreisel, S. Zhong, Y. Li, B. M. Andersen, P. Hirschfeld, J. Wang, Orbital-selective high-temperature Cooper pairing developed in the two-dimensional limit. *Nano Lett.* **22**, 3245–3251 (2022).

26. R. M. Fernandes, A. I. Coldea, H. Ding, I. R. Fisher, P. J. Hirschfeld, G. Kotliar, Iron pnictides and chalcogenides: a new paradigm for superconductivity. *Nature* **601**, 35–44 (2022).

27. Q. Si, R. Yu, E. Abrahams, High-temperature superconductivity in iron pnictides and chalcogenides. *Nat. Rev. Mater*, **1**, 1-15 (2016).

28. S. Lafuerza, H. Gretarsson, F. Hardy, T. Wolf, C. Meingast, G. Giovannetti, M. Capone, A. S. Sefat, Y.-J. Kim, P. Glatzel, L. de' Medici, Evidence of Mott physics in iron pnictides from x-ray spectroscopy. *Phys. Rev. B* **96**, 045133 (2017).

29. A. Kreisel, P. J. Hirschfeld, B. M. Andersen, Theory of Spin-Excitation Anisotropy in the Nematic Phase of FeSe Obtained From RIXS Measurements. *Frontiers in Physics*, **10**, 859424 (2022)

30. T. M. McQueen, A. J. Williams, P. W. Stephens, J. Tao, Y. Zhu, V. Ksenofontov, F. Casper, C. Felser, R. J. Cava, Tetragonal-to-orthorhombic structural phase transition at 90 K in the superconductor $Fe_{1.01}$Se. *Phys. Rev. Lett.* **103**, 057002 (2009).

31. J. K. Glasbrenner, I. I. Mazin, H. O. Jeschke, P. J. Hirschfeld, R. M. Fernandes, R. Valentí, Effect of magnetic frustration on nematicity and superconductivity in iron chalcogenides. *Nat. Phys.* **11**, 8 (2015).

32. A. Kreisel, P. J. Hirschfeld, B. M. Andersen, On the Remarkable Superconductivity of FeSe and Its Close Cousins. *Symmetry* **12**, 1402 (2020).

33. F. Wang, S. A. Kivelson, D.-H. Lee, Nematicity and quantum paramagnetism in FeSe. *Nature Phys* **11**, 959–963 (2015).

34. J. P. Sun, K. Matsuura, G. Z. Ye, Y. Mizukami, M. Shimozawa, K. Matsubayashi, M. Yamashita, T. Watashige, S. Kasahara, Y. Matsuda, J.-Q. Yan, B. C. Sales, Y. Uwatoko, J.-G. Cheng, T. Shibauchi, Dome-shaped magnetic order competing with high-temperature superconductivity at high pressures in FeSe. *Nat Commun* **7**, 12146 (2016).

35. J. H. J. Martiny, A. Kreisel, B. M. Andersen, Theoretical study of impurity-induced magnetism in FeSe. *Phys. Rev. B* **99**, 014509 (2019).

36. S. Y. Song, J. H. J. Martiny, A. Kreisel, B. M. Andersen, J. Seo, Visualization of local magnetic moments emerging from impurities in Hund's metal states of FeSe. *Phys. Rev. Lett.* **124**, 117001 (2020).

37. S. Y. Song, J. Seo, Local magnetism induced by non-magnetic impurities in FeSe in proximity to s-wave superconductivity. *Appl. Phys. Lett.* **119**, 052601 (2021).

38. W. Li, Y. Zhang, P. Deng, Z. Xu, S.-K. Mo, M. Yi, H. Ding, M. Hashimoto, R. G. Moore, D.-H. Lu, X. Chen, Z.-X. Shen, Q.-K. Xue, Stripes developed at the strong limit of nematicity in FeSe film. *Nature Phys* **13**, 957–961 (2017).

39. M. N. Gastiasoro, P. J. Hirschfeld, B. M. Andersen, Impurity states and cooperative magnetic order in Fe-based superconductors. *Phys. Rev. B* **88**, 220509 (2013).

40. M. N. Gastiasoro, B. M. Andersen, Local Magnetization Nucleated by Non-magnetic Impurities in Fe-based Superconductors. *J Supercond Nov Magn* **28**, 1321–1324 (2015).

41. Y. Suzuki, T. Shimojima, T. Sonobe, A. Nakamura, M. Sakano, H. Tsuji, J. Omachi, K. Yoshioka, M. Kuwata-Gonokami, T. Watashige, R. Kobayashi, S. Kasahara, T. Shibauchi, Y. Matsuda, Y. Yamakawa, H. Kontani, K. Ishizaka, Momentum-dependent sign inversion of orbital order in superconducting FeSe. *Phys. Rev. B* **92**, 205117 (2015).

42. M. D. Watson, T. K. Kim, L. C. Rhodes, M. Eschrig, M. Hoesch, A. A. Haghighirad, A. I. Coldea, Evidence for unidirectional nematic bond ordering in FeSe. *Phys. Rev. B* **94**, 201107 (2016).

43. D. Liu, C. Li, J. Huang, B. Lei, L. Wang, X. Wu, B. Shen, Q. Gao, Y. Zhang, X. Liu, Y. Hu, Y. Xu, A. Liang, J. Liu, P. Ai, L. Zhao, S. He, L. Yu, G. Liu, Y. Mao, X. Dong, X. Jia, F. Zhang, S. Zhang, F. Yang, Z. Wang, Q. Peng, Y. Shi, J. Hu, T. Xiang, X. Chen, Z. Xu, C. Chen, X. J. Zhou, Orbital origin of extremely anisotropic superconducting gap in nematic phase of FeSe superconductor. *Phys. Rev. X* **8**, 031033 (2018).

44.     T. Terashima, N. Kikugawa, A. Kiswandhi, E.-S. Choi, J. S. Brooks, S. Kasahara, T. Watashige, H. Ikeda, T. Shibauchi, Y. Matsuda, T. Wolf, A. E. Böhmer, F. Hardy, C. Meingast, H. v. Löhneysen, M.-T. Suzuki, R. Arita, S. Uji, Anomalous Fermi surface in FeSe seen by Shubnikov-de Haas oscillation measurements. *Phys. Rev. B* **90**, 144517 (2014).

45.     M. D. Watson, T. K. Kim, A. A. Haghighirad, N. R. Davies, A. McCollam, A. Narayanan, S. F. Blake, Y. L. Chen, S. Ghannadzadeh, A. J. Schofield, M. Hoesch, C. Meingast, T. Wolf, A. I. Coldea, Emergence of the nematic electronic state in FeSe. *Phys. Rev. B* **91**, 155106 (2015).

46.     A. Kostin, P. O. Sprau, A. Kreisel, Y. X. Chong, A. E. Böhmer, P. C. Canfield, P. J. Hirschfeld, B. M. Andersen, J. C. S. Davis, Imaging orbital-selective quasiparticles in the Hund's metal state of FeSe. *Nature Mater* **17**, 869–874 (2018).

47.     H. Kim, Y. Hasegawa, Site-dependent evolution of electrical conductance from tunneling to atomic point contact. *Phys. Rev. Lett.* **114**, 206801 (2015).

48.     A. Yazdani, B. A. Jones, C. P. Lutz, M. F. Crommie, D. M. Eigler, Probing the local effects of magnetic impurities on superconductivity. *Science* **275**, 1767–1770 (1997).

49.     A. V. Balatsky, I. Vekhter, J.-X. Zhu, Impurity-induced states in conventional and unconventional superconductors. *Rev. Mod. Phys.* **78**, 373–433 (2006).

50.     K. J. Franke, G. Schulze, J. I. Pascual, Competition of superconducting phenomena and Kondo screening at the nanoscale. *Science* **332**, 940–944 (2011).

51.     S. H. Pan, E. W. Hudson, J. C. Davis, Vacuum tunneling of superconducting quasiparticles from atomically sharp scanning tunneling microscope tips. *Appl. Phys. Lett.* **73**, 2992–2994 (1998).

52.     C. Li, X. Wu, L. Wang, D. Liu, Y. Cai, Y. Wang, Q. Gao, C. Song, J. Huang, C. Dong, J. Liu, P. Ai, H. Luo, C. Yin, G. Liu, Y. Huang, Q. Wang, X. Jia, F. Zhang, S. Zhang, F. Yang, Z. Wang, Q. Peng, Z. Xu, Y. Shi, J. Hu, T. Xiang, L. Zhao, X. J. Zhou, Spectroscopic evidence for an additional symmetry breaking in the nematic state of FeSe superconductor. *Phys. Rev. X* **10**, 031033 (2020).

53.     L. Benfatto, L. Fanfarillo, Nematic pairing from orbital-selective spin fluctuations in FeSe. *npj Quant Mater* **3**, 1–7 (2018).

54.     A. V. Balatsky, I. Vekhter, J.-X. Zhu, Impurity-induced states in conventional and unconventional superconductors. *Rev. Mod. Phys*. **78**, 373 (2006).

55.     T. Watashige, Y. Tsutsumi, T. Hanaguri, Y. Kohsaka, S. Kasahara, A. Furusaki, M. Sigrist, C. Meingast, T. Wolf, H. v. Löhneysen, T. Shibauchi, and Y. Matsuda, Evidence for Time-Reversal Symmetry Breaking of the Superconducting State near Twin-Boundary Interfaces in FeSe Revealed by Scanning Tunneling Spectroscopy. *Phys. Rev. X* **5**, 031022 (2015)

56.     Z Ge, C Yan, H Zhang, D Agterberg, M Weinert, L Li, Evidence for d-wave superconductivity in single Layer FeSe/SrTiO3 probed by quasiparticle scattering off step edges. *Nano Lett.* **19**, 2497–2502 (2019)

57.     L. Shan, Y.-L. Wang, J. Gong, B. Shen, Y. Huang, H. Yang, C. Ren, H.-H. Wen, Evidence of multiple nodeless energy gaps in superconducting $Ba_{0.6}K_{0.4}Fe_2As_2$ single crystals from scanning tunneling spectroscopy. *Phys. Rev. B* **83**, 060510 (2011).

58.     C.-L. Song, Y.-L. Wang, Y.-P. Jiang, Z. Li, L. Wang, K. He, X. Chen, J. E. Hoffman, X.-C. Ma, Q.-K. Xue, Imaging the Electron-Boson coupling in superconducting FeSe films using a scanning tunneling microscope. *Phys. Rev. Lett.* **112**, 057002 (2014).

59.     C. M. Yim, C. Trainer, R. Aluru, S. Chi, W. N. Hardy, R. Liang, D. Bonn, P. Wahl, Discovery of a strain-stabilised smectic electronic order in LiFeAs. *Nat Commun* **9**, 2602 (2018).

60.     J.-X. Yin, S. S. Zhang, G. Dai, Y. Zhao, A. Kreisel, G. Macam, X. Wu, H. Miao, Z.-Q. Huang, J. H. J. Martiny, B. M. Andersen, N. Shumiya, D. Multer, M. Litskevich, Z. Cheng, X. Yang, T. A. Cochran, G. Chang, I. Belopolski, L. Xing, X. Wang, Y. Gao, F.-C. Chuang, H. Lin, Z. Wang, C. Jin, Y. Bang, M. Z. Hasan, Quantum phase transition of correlated iron-based superconductivity in $LiFe_{1-x}Co_xAs$. *Phys. Rev. Lett.* **123**, 217004 (2019).

61.     L. Cao, W. Liu, G. Li, G. Dai, Q. Zheng, Y. Wang, K. Jiang, S. Zhu, L. Huang, L. Kong, F. Yang, X. Wang, W. Zhou, X. Lin, J. Hu, C. Jin, H. Ding, H.-J. Gao, Two distinct superconducting states controlled by orientations of local wrinkles in LiFeAs. *Nat Commun* **12**, 6312 (2021).

**Experimental methods**

We have conducted the experiment using a SPECS Joule-Thompson STM at 1.2 K under a base pressure of < $10^{-10}$ mbar. A SPECS-Nanonis controller was employed. Freshly cleaved surfaces were obtained by delaminating a small sample of FeSe in low vacuum of < 10-7 mbar, followed by rapid sample transfer to the cryogenic chamber. The dI/dV spectra and differential conductance maps of superconducting gaps were obtained using a lock-in technique with the modulation frequency 785 Hz.

**Acknowledgments**

This work was primarily supported by the U.S. Department of Energy, Office of Science, Basic Energy Sciences, Materials Sciences and Engineering Division. Scanning tunneling microscopy was performed as a user project at the Center for Nanophase Materials Sciences (CNMS), which is a US Department of Energy, Office of Science User Facility at Oak Ridge National Laboratory. S.Y.S. received additional support for data analysis from the National Research Foundation of Korea (NRF) grant funded by the Korea government (MSIT) (RS-2024-00337725).

# Supplementary materials

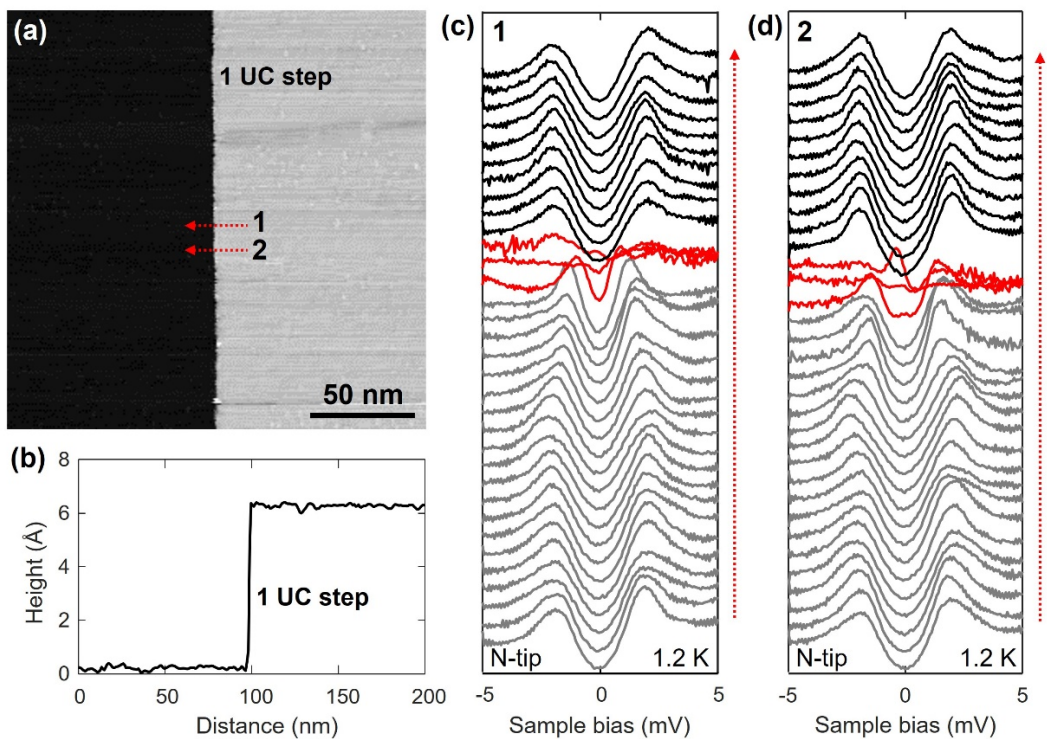

**Figure S1** In-gap states at the step edge of FeSe. (a) Topographic image of one-unit cell (1 UC) step edge FeSe. (b) Topographic line profile across the 1 UC step. (c, b) Spatial variation of superconducting gaps across the step edge (red dotted arrow in (a)). Clear in-gap states appear at the step edges (red spectra).

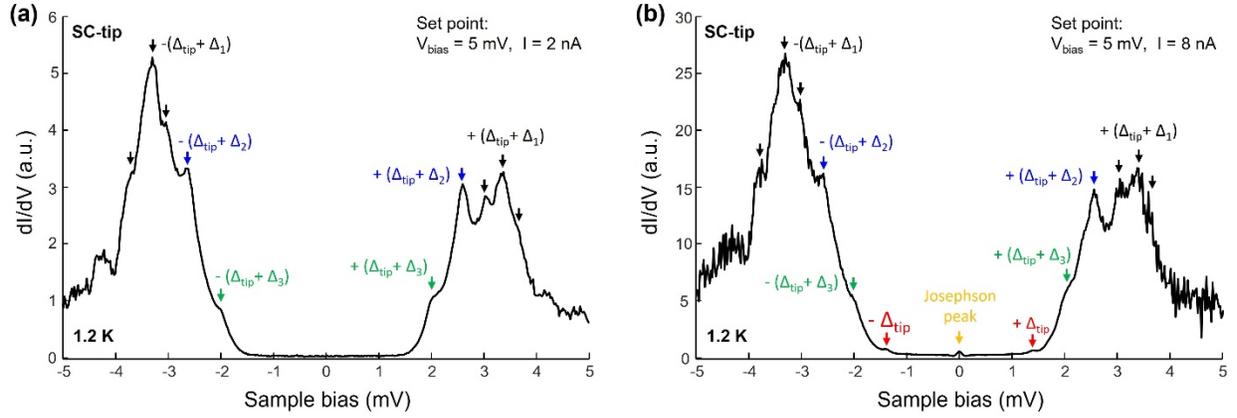

**Figure S2** High energy resolution superconducting gap on FeSe obtained by a Pb-coated superconducting tip (Setpoint: $V_{bias}$ = 5 mV, I = 8 nA). Red and the blue arrows indicate the positions of Andreev resonance peaks ($\Delta_{tip}$ = ± 1.374mV), and a Josephson peak (0 mV), respectively. The black arrows (± ($\Delta_{tip}$ + $\Delta_1$) = ± 3.80, ± 3.30, and ± 3.02 meV), blue arrows (± ($\Delta_{tip}$ + $\Delta_2$) = ± 2.63 meV), and green arrows (± ($\Delta_{tip}$ + $\Delta_3$) = ± 2.03 meV) originate from the coherence peaks of gap $\Delta_1$ (± 2.42, ± 1.93, and ± 1.65 meV), $\Delta_2$ (± 1.25 meV), and $\Delta_3$ (± 0.66 meV), respectively.

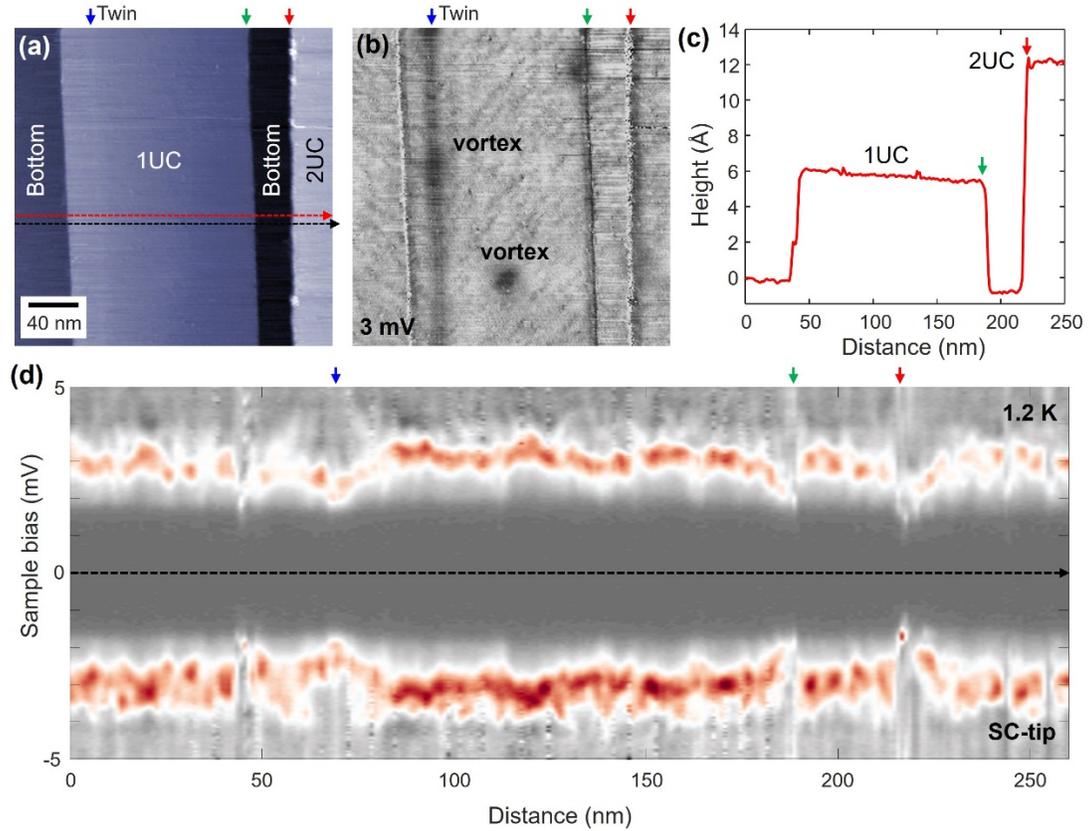

**Figure S3 (a)** STM image of steps (green, and red arrows) and a twin boundary (blue arrow) on FeSe ($V_{bias}$ = 100 mV, I = 100pA). **(b)** Differential conductance map of steps and the twin boundary at 3 mV (I = 100pA). **(c)** Topographic line profile across the 1-unit cell (UC) step and 2 UC step (red dashed line in (a)). **(d)** Spatial variation of superconducting gaps of FeSe across the steps and the twin boundary (black dashed arrow in (a)) obtained using a Pb-coated superconducting tip at 1.2 K.

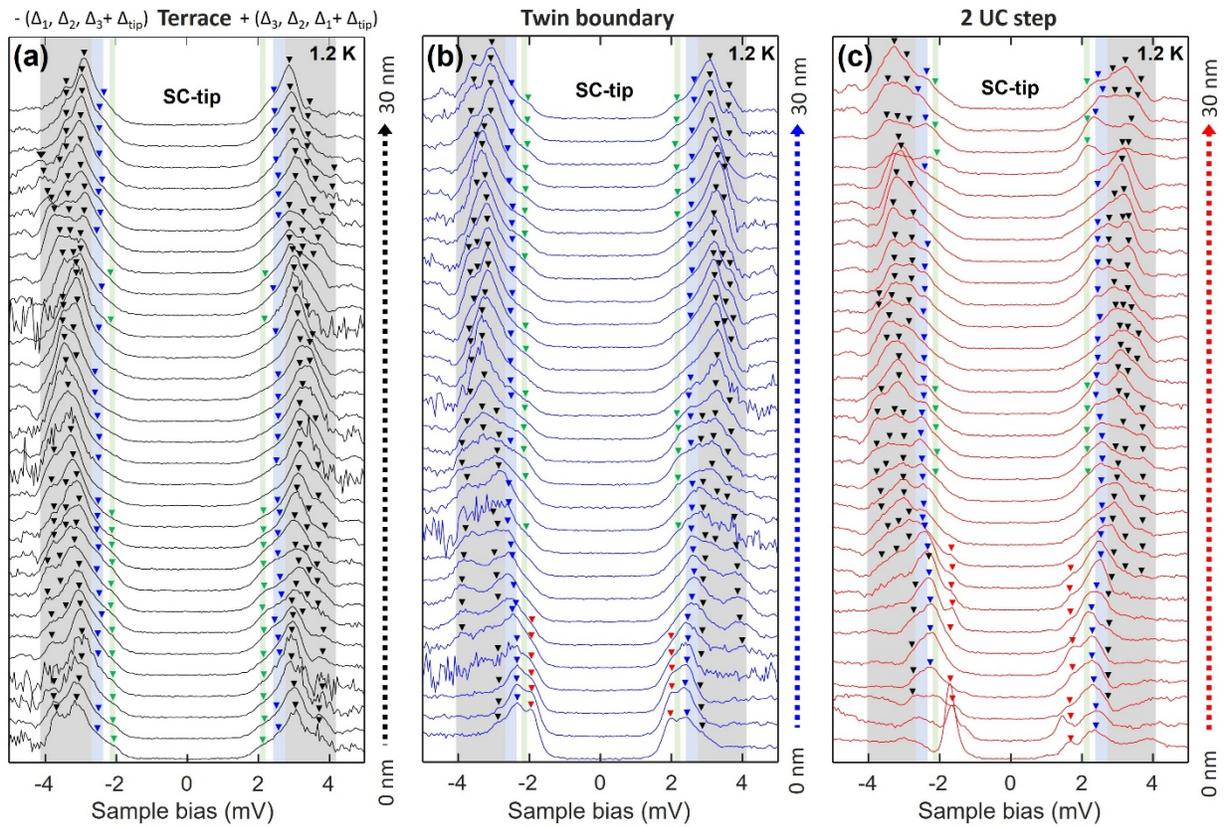

**Figure S4 (a-c)** Spatial variations of superconducting coherence peaks across the terrace, the twin boundary, and step (black, blue, and red dashed arrows in Fig.4a in the main text) obtained using a Pb-coated superconducting tip at 1.2 K. The black, blue, green, and red triangles indicate the peak positions of the gap $\Delta_1$, gap $\Delta_2$, gap $\Delta_3$, and in-gap state, respectively.

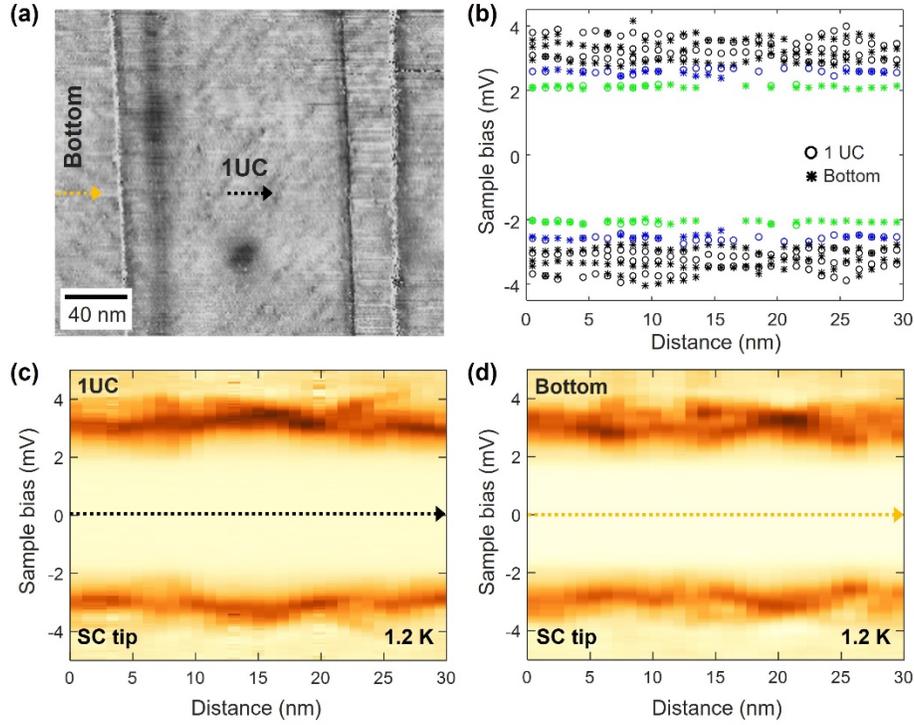

**Figure S5 Spatial variations of superconducting gaps on the bottom and the 1 UC step. (a)** Differential conductance map of steps and the twin boundary at 3 mV. **(b)** The estimated trajectories of superconducting gaps on the bottom (circles), and one unit cell (1 UC) step (stars) on FeSe. **(c, d)** Spatial variation of the superconducting gaps on Bottom and 1 UC step.

| **Normal tip** (at 1.2K) | | |
|---|---|---|
| $\Delta_1$ | $\Delta_2$ | $\Delta_3$ |
| ± 2 ~ 2.4 mV | ± 1 ~ 1.2 mV | - |
| **Pb coated superconducting tip** ($\Delta_{tip}$ = 1.374 mV) (at 1.2K) | | |
| $\Delta_1$ ($\Delta_{tip}+\Delta_1$) | $\Delta_2$ ($\Delta_{tip}+\Delta_2$) | $\Delta_3$ ($\Delta_{tip}+\Delta_3$) |
| ± 1.314 ~ 2.666 mV (± 2.688~ 4.040 mV) | ± 0.963 ~ 1.365 mV (± 2.337~ 2.739 mV) | ± 0.661 ~ 0.812 mV (± 2.035~ 2.186 mV) |

**Table S1** Energy positions of superconducting coherence peaks on the terrace of FeSe by using a normal tip and a Pb-coated superconducting tip.

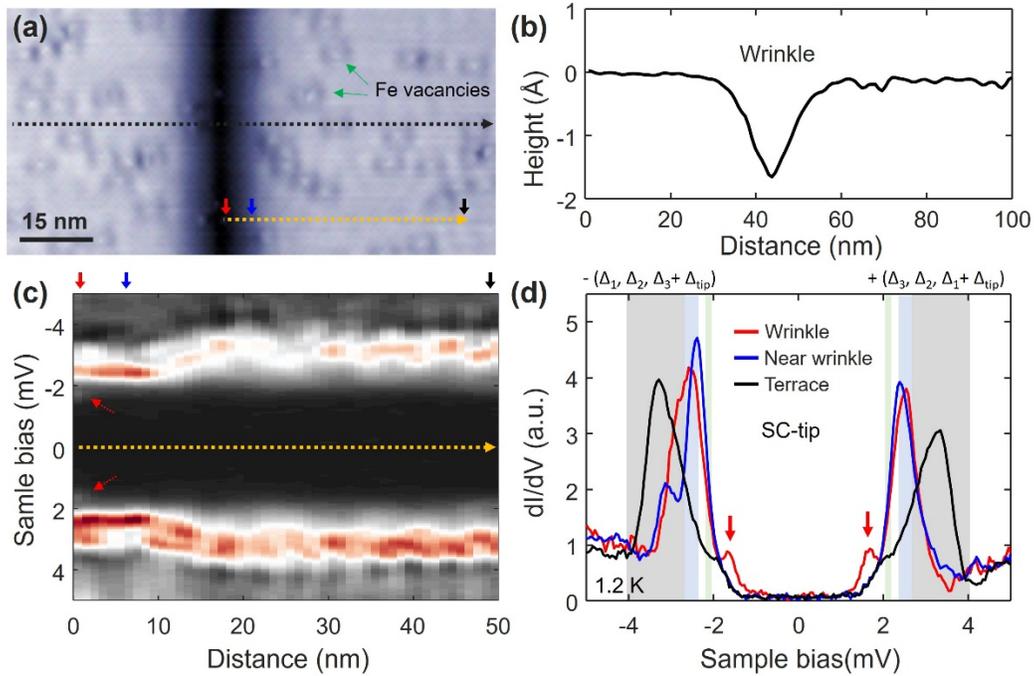

**Figure. S6 In-gap state near a wrinkle (a)** Topographic image of wrinkle. **(b)** The topographic line profile across the wrinkle (black dotted line in (a)). **(c)** Variation of superconducting gaps along the orange dotted line near wrinkle in (a). The red dotted arrows indicate the in-gap state at the wrinkle. **(d)** The dI/dV spectra obtained at the wrinkle (red arrow in (a)), near the wrinkle (blue arrow in (a)), and on the terrace (black arrow in (a)). The gray, blue, and green box indicate the energy ranges of the gap $\Delta_1$, gap $\Delta_2$, and gap $\Delta_3$, respectively.

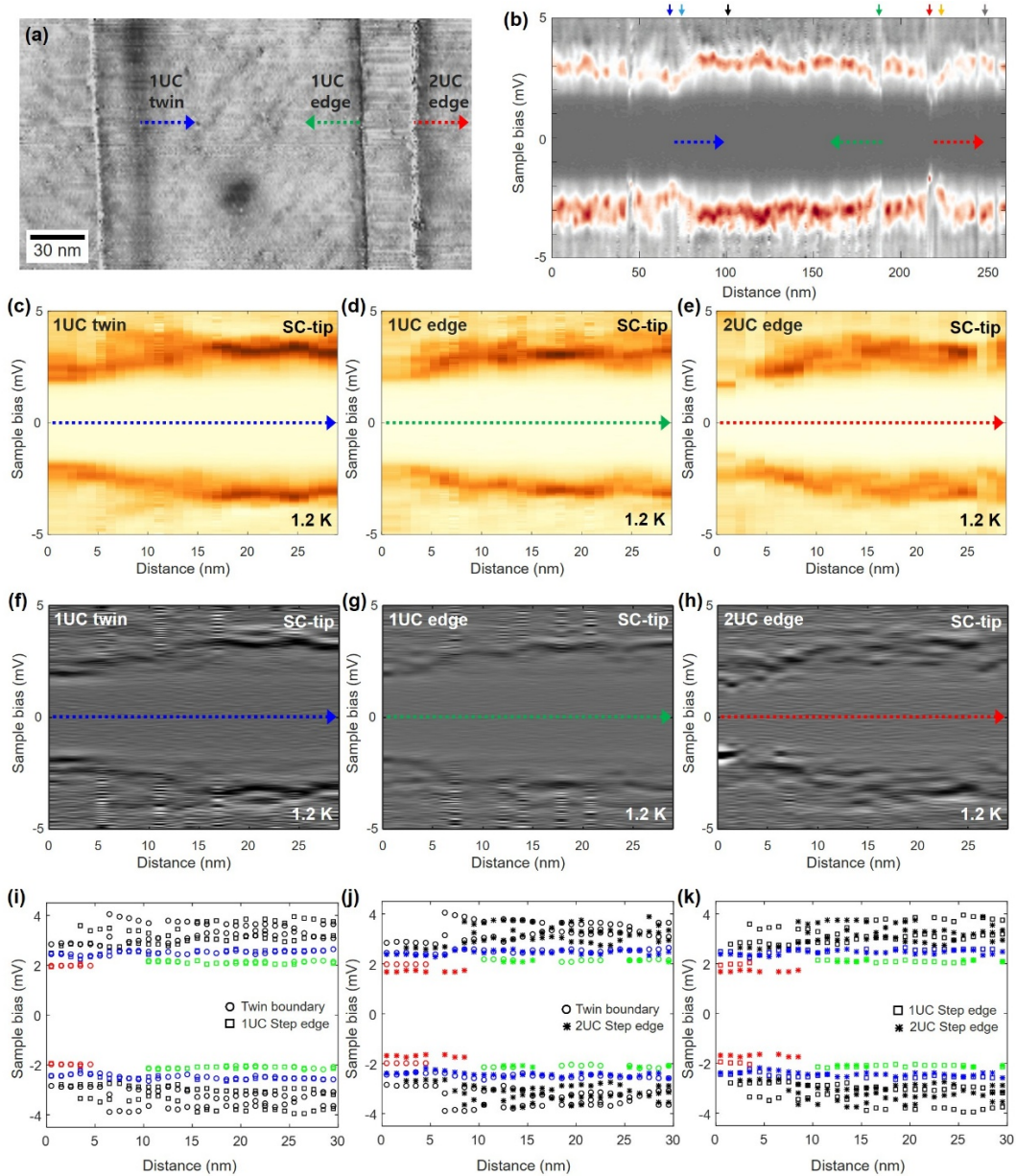

**Figure S7 Comparison spatial variations of superconducting gaps on the bottom and the 1UC step. (a)** Differential conductance map of steps and the twin boundary at 3 mV. **(b)** Spatial variation of superconducting gaps across the twin boundary and step edges. **(c - e)** Detailed gap structure near the twin boundary (blue dotted arrow in (a)), 1 unit cell (UC) step (green dotted arrow in (a)), and 2UC step (red arrow in (a)). **(f - h)** Second derivative curves near the twin boundary (blue dotted arrow in (a)), 1 unit cell (UC) step (green dotted arrow in (a)), and 2UC step (red arrow in (a)). **(i)** Comparison of trajectories of superconducting gaps between near the twin boundary (circles), and 1UC step edge (squares). **(j)** Comparison of trajectories of superconducting gaps between near the twin boundary (circles), and 2UC step edge (stars). **(k)** Comparison of trajectories of superconducting gaps between near the 1UC steps (squares), and 2UC step edge (stars).

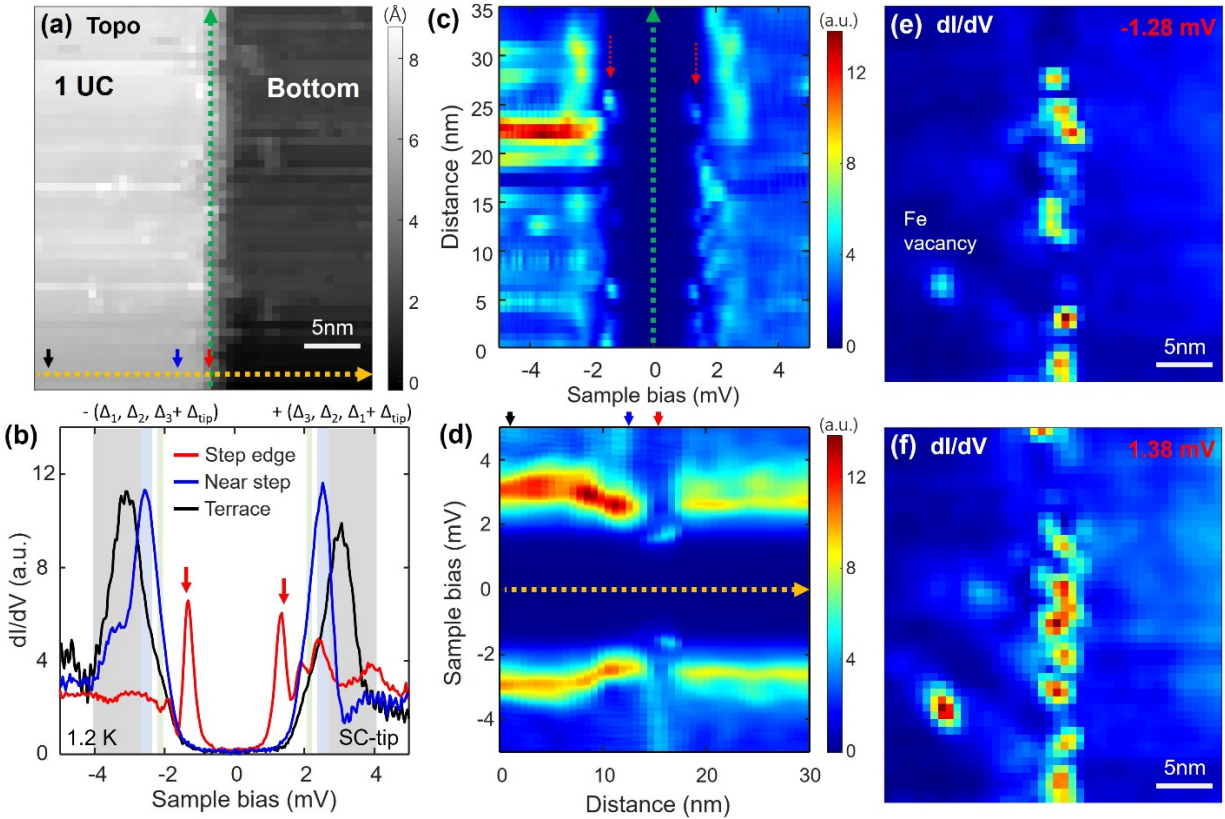

**Figure S8** Spatial distribution of in-gap states near the step edge of FeSe. (a) Topographic image of one-unit cell step on FeSe. (b) The dI/dV spectra obtained at step edge (red arrow in (a)), near the step edge (blue arrow in (a)), and on the terrace (blue arrow in (a)). (c) Variation of superconducting gaps across the step edge (orange dotted arrow in (a)). (d) Variation of in-gap state along the step edge (green dotted arrow in (a)). (e, f) The dI/dV maps at -1.28 mV and 1.38 mV which correspond to the energy position of in-gap states (red arrows in (b)). The dI/dV maps show the localized distributions of in-gap states near the step edge.